\documentclass[12pt]{iopart}
\usepackage{iopams}  

\usepackage{graphicx}
\usepackage{dcolumn}
\usepackage{bm}
\usepackage{subfigure}
\usepackage{cite}
\usepackage{enumerate}
\usepackage{rotating}

\newcommand{\ion}[2]{\mbox{\ensuremath{^{#2}}#1\ensuremath{^+}}}

\newcommand{\ket}[1]{\ensuremath{\bigl| #1 \bigr>}}

\newcommand{\lev}[2]{\mbox{#1$_{\mbox{\tiny$#2$}}$}}

\newcommand{\Ca}[1]{\ion{Ca}{#1}}
\newcommand{\Mg}[1]{\ion{Mg}{#1}}
\newcommand{\Be}[1]{\ion{Be}{#1}}
\newcommand{\Sr}[1]{\ion{Sr}{#1}}
\newcommand{\Ba}[1]{\ion{Ba}{#1}}

\begin{document}

\title[Dark-resonance Doppler cooling and high fluorescence in trapped Ca-43 ions]{Dark-resonance Doppler cooling and\, \\high fluorescence in trapped Ca-43 ions\, \\at intermediate magnetic field}

\author{D. T. C. Allcock\footnote{Current address: National Institute of Standards and Technology, 325 Broadway, Boulder, CO 80305, USA}, T. P. Harty, M. A. Sepiol, H. A. Janacek, \\ C. J. Ballance, A. M. Steane, D. M. Lucas, D. N. Stacey}

\address{Department of Physics, University of Oxford, Clarendon Laboratory, Parks Road, Oxford OX1 3PU, United Kingdom}

\ead{d.lucas@physics.ox.ac.uk}

\begin{abstract}
We demonstrate simple and robust methods for Doppler cooling and obtaining high fluorescence from trapped $^{43}$Ca$^+$ ions at a magnetic field of 146\,Gauss. This field gives access to a magnetic-field-independent ``atomic clock'' qubit transition within the ground level hyperfine structure of the ion, but also causes the complex internal structure of the 64 states relevant to Doppler cooling to be spread over many times the atomic transition line-width. Using a time-dependent optical Bloch equation simulation of the system we develop a simple scheme to Doppler-cool the ion on a two-photon dark resonance, which is robust to typical experimental variations in laser intensities, detunings and polarizations. We experimentally demonstrate cooling to a temperature of 0.3\,mK, slightly below the Doppler limit for the corresponding two-level system, and then use Raman sideband laser cooling to cool further to the ground states of the ion's radial motional modes. These methods will enable two-qubit entangling gates with this ion, which is one of the most promising qubits so far developed.
\end{abstract}



\maketitle

\section{Introduction}
\label{introduction}

Trapped ion qubits based on magnetic field-independent ``atomic clock'' transitions between hyperfine states are well established in quantum information processing.  It is advantageous to use a clock transition which occurs at a non-zero magnetic field (i.e., in the intermediate field regime) as this lifts the degeneracy with other magnetic sublevels and provides a well-defined quantization axis for the ion.  Previously, such intermediate field clock qubits have only been demonstrated in \Be{9} \cite{Langer05} and \Mg{25} \cite{Ospelkaus11}.  Because these ions lack low-lying D levels, they have a closed two-level Doppler cooling transition (\lev{S}{1/2}~$-$~\lev{P}{3/2}) which makes cooling them at any magnetic field straightforward.  In this paper we describe methods to overcome the complication of cooling and obtaining high fluorescence from ions with low-lying D levels such as \Ca{43}, \Sr{87}, and \Ba{135, 137} at intermediate field.  These ions offer the advantage of high-fidelity readout through electron shelving \cite{Myerson08} and transitions that are conveniently located at diode laser-accessible wavelengths in the visible and near-IR rather than in the UV.\\

We make use of a Doppler-cooling scheme to cool singly-ionized $^{43}$Ca in a Paul trap to a temperature of 0.3\,mK in an external field of 146 Gauss at which the first intermediate field clock transition occurs.  In recent work \cite{Harty14} on this system we have achieved a coherence time of 50(10)\,s (the longest ever recorded for a single qubit), demonstrated a combined state preparation and read-out error of  $6.8(6)\times10^{-4}$, and performed single-qubit logic operations with average errors of $1.0(3)\times10^{-6}$ (representing more than an order of magnitude improvement on any previous work).  These operations, however, do not directly involve the ion's motional state.  In order to drive two-qubit logic operations we need control over motional as well as internal states of the ion.  We have therefore developed a Doppler cooling scheme which leaves an ion confined to the lowest few motional states of the trap.  We also demonstrate that at this point Raman sideband cooling can be applied to the ion to reach the ground state.  The laser system needed to achieve the Doppler cooling has also to be readily adaptable to produce high fluorescence; this is necessary since our low figure for the state read-out error depends on a high photon count rate.\\

The usual method for Doppler cooling calcium ions involves laser cooling on the 4\lev{S}{1/2}~$-$~4\lev{P}{1/2} transition at 397\,nm, with a repumper at 866\,nm to re-excite from the metastable 3\lev{D}{3/2} level.  This has been shown to work well on \Ca{40} and even on \Ca{43} in an external field of a few Gauss \cite{Lucas07, Benhelm08}.  However, \Ca{43} in a field of 146 Gauss presents difficulties.  The population can spread over 64 states since there is no ``closed set" to which the ion can be confined.  The ground level hyperfine splitting is 3.2\,GHz, and the external field causes adjacent states in this manifold to be split by $\sim$50\,MHz, larger than the 4\lev{S}{1/2}~$-$~4\lev{P}{1/2} natural line-width of 22.3\,MHz.  The laser radiation applied must be able to access all states, and we still wish to cool to a temperature comparable with that which would be reached by a simple two-state system.  We achieve this by exploiting the properties of a dark resonance and using only 3 laser frequencies. Furthermore, the method is not highly sensitive to the frequencies, polarizations or intensities of the beams.  Enhanced cooling close to such a resonance has long been predicted \cite{Reiss96, Reiss02} and recently experimentally demonstrated \cite{Harlander12, Rossnagel15} in simpler 8-state systems without hyperfine structure (\Ca{40}) at low magnetic field.  However this cooling scheme is of limited utility in systems such as these in which the Doppler limit is straightforward to obtain by conventional Doppler cooling.\\

In section~\ref{coolingscheme}, we explain the scheme and how the temperature attainable depends on the choice of experimental parameters.  In section~\ref{experimental}, we describe how the scheme was realized in practice and give the experimental results.  This section also shows that with the addition of a Raman sideband cooling step the ion can be cooled close to the motional ground state of the trap.

\section{The dark-resonance cooling scheme}
\label{coolingscheme}

\begin{figure}
 \centering
 \includegraphics[width=0.60\textwidth]{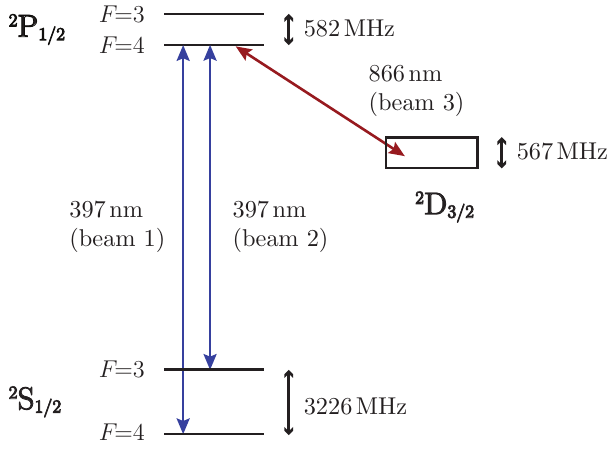}
 \caption{Energy levels 4\lev{S}{1/2}, 4\lev{P}{1/2} and 3\lev{D}{3/2} of \Ca{43} in zero field (hyperfine structure in 3\lev{D}{3/2} not shown).  At 146 Gauss the S and P states are still in well-defined groups corresponding to the quantum number $F$.  Mixing is much greater in the D states.  Two 397\,nm laser beams excite the ion from the ground level, while a beam at 866\,nm acts as a repumper from 3\lev{D}{3/2}, mainly to 4\lev{P}{1/2} $F=4$.}
 \label{Ca43levs}
\end{figure}

Figure~\ref{Ca43levs} shows the relevant energy levels of \Ca{43} (4\lev{S}{1/2}, 4\lev{P}{1/2} and 3\lev{D}{3/2}) in zero field.   Even in the presence of a field of 146 Gauss, the S and P states are still in well-defined groups corresponding to the quantum number $F$, so it is convenient to distinguish them with this quantity.  This is not true of the 32 D states, but fortunately we do not need to consider these in any detail; they are shown as a block in the figure.  Because of the large ground level hyperfine splitting, two 397\,nm laser beams are required to excite the ion efficiently from the ground level, one taking it from $F=4$ (beam 1), the other from $F=3$ (beam 2).  The frequencies are tuned close to the $F=4$ level of 4\lev{P}{1/2}, though off-resonant excitation can occur to $F=3$.  A single intense beam at 866\,nm (beam 3) acts as a repumper from 3\lev{D}{3/2}, again primarily to $F=4$ of 4\lev{P}{1/2}.  The Einstein $A$-coefficients for the \lev{P}{1/2}$\to$\lev{S}{1/2} and \lev{P}{1/2}$\to$\lev{D}{3/2} decays are $1.32\times10^8$\,s$^{-1}$ and $8.4\times10^6$\,s$^{-1}$ respectively, so the ion undergoes many \lev{S}{1/2}$\leftrightarrow$\lev{P}{1/2} cycles for every decay to the metastable \lev{D}{3/2} level.\\

To provide a benchmark against which our cooling scheme may be judged we first model the system as a two-state ion which undergoes simple harmonic motion in an isotropic trap, subject to a single red-detuned laser beam at 397\,nm.  We thus neglect decays to 3\lev{D}{3/2} entirely, also the degeneracies of the S and P levels.  The theory for this situation is well-known~\cite{Leibfried03}.  Cooling occurs due to the preferential absorption of photons when the ion is moving towards the laser source, and an equilibrium temperature is reached when the cooling is balanced by the heating effects associated with spontaneous emission and the stochastic nature of the absorption process.  We neglect heating processes other than these.  Let the light be of wavelength $\lambda$ and propagating in the positive $x$-direction.  We consider the velocity component $V$ of the ion in the positive $x$-direction.  We define the detuning $\Delta\nu$ of the laser from resonance to be positive if it is towards higher frequency.  Let $R(\Delta \nu)$ be the rate at which photons are absorbed by a stationary ion, and $R'(\Delta \nu)$ be the derivative with respect to $V$ at $V = 0$.   Then the equilibrium temperature $T$ is given by

\begin{equation}\label{equibtempeqn}
\frac{k_BT}{h}=-\frac{1}{\lambda}\frac{R(\Delta \nu)}{R'(\Delta \nu)}.
\end{equation}

The minimum temperature $T_{min}$ occurs when the transition is unsaturated, at 
${\Delta \nu = -A/4\pi}$, i.e. a red detuning of 10.5\,MHz.  We then have ${T_{min}=A\hbar/2k_B \approx 0.5}$\,mK.  We show in the following that our cooling scheme makes temperatures of this order feasible in high-field \Ca{43}, despite the complex level structure.\\

\begin{figure}
 \centering
 \includegraphics[width=0.75\textwidth]{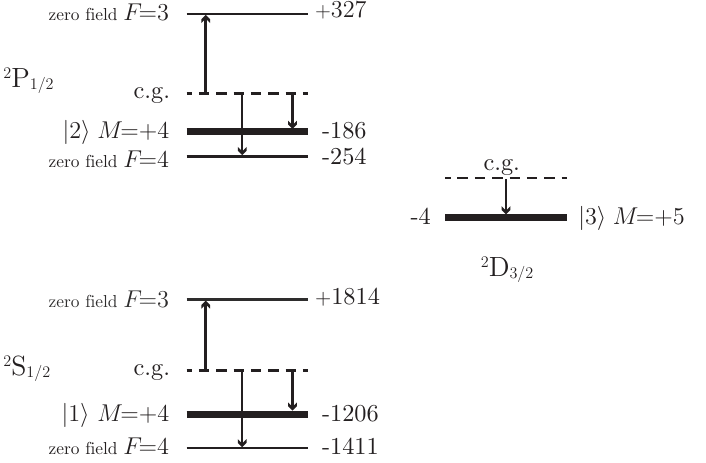}
 \caption{The states relevant to the cooling scheme (see text).  The diagram is not to scale.  Separations are given in MHz with respect to the positions of the centres of gravity (c.g.) of the levels.  The positions of the zero-field hyperfine levels are also shown.}
 \label{Ca43lambda}
\end{figure}

We return to the true level structure of \Ca{43}.  Equation~\ref{equibtempeqn} still applies, but the photon absorption rate and its derivative with respect to $V$ now depend on all three laser detunings; we label them simply $R$ and $R'$.  The reason we are able to cool this ion effectively is that in the vicinity of a dark resonance we can combine a high value of $R'$ with a low value of $R$.  The three states involved in the particular dark resonance we use are shown in figure~\ref{Ca43lambda}; they are \ket{F = 4, M = +4} of the ground level, \ket{F = 4, M = +4} of the P level and \ket{M = +5} of the D level.  We refer to them as \ket{1}, \ket{2} and \ket{3} respectively.  The dark resonance occurs when the detuning from resonance of the laser L3 exciting the transition \ket{3} to \ket{2} is equal to the detuning from resonance of the laser L1 exciting from \ket{1} to \ket{2}.   The reason for choosing this resonance is that with suitably polarized beams it is completely dark; that is, there is no route out of \ket{1} or \ket{3} except to \ket{2}.  When the above frequency condition is fulfilled, therefore, the population of the entire P manifold drops sharply to zero.  In practice, this drop is limited by the motion of the ion and the finite laser line-widths but is still rapid enough to give a large value of $R'$.  Red detuning of L3 from the bottom of the dark resonance by a few MHz gives the conditions for efficient cooling.  The highest values of $R'$, and hence the lowest temperatures, are obtained with L1 and L3 counter-propagating, since then the Doppler shifts are in opposite senses.\\

The polarization requirements are set by the fact that $\sigma^-$ polarization should not be present in L1, and that the ion must also be able to reach \ket{2} by laser excitation from all the S and D states in the system, though not necessarily of course in a single step.  There is a second such resonance involving the corresponding states with the signs of $M$ reversed.\\

\subsection{Choice of experimental conditions}
\label{experimentalconditions}

Within the constraints listed above, there remains wide scope for choice of experimental conditions.   Our objective was to arrive at a set which was convenient in practice, and would quickly and reliably lead to temperatures comparable with or lower than the Doppler limit given above.  This is enough to allow further cooling by other techniques to be straightforward.  Robustness against slight changes in conditions is much more important than achieving the lowest possible temperature.  It is also necessary that the system should be easily adaptable to give a high level of fluorescence, since this is required for qubit read-out.  Our read-out method is described in \cite{Harty14}; to distinguish which of the clock states the ion occupies we apply microwave and laser pulses which transfer the ion to the $3^2$\lev{D}{5/2} level only if it starts in a particular one of them.  The  $3^2$\lev{D}{5/2} level is a shelf which is outside the cooling cycle so the presence or absence of fluorescence when this cycle is resumed provides the required discrimination.  For the level of fluorescence not to be a limiting factor in this process, we required a count-rate which, given the efficiency of our detection system, translates to a need for the total population in the \lev{P}{1/2} level to be of the order of 10\%.  This would give a value of $R$ much too high for adequate cooling.  We therefore need a parameter set which can readily be ``switched'' between cooling and high fluorescence modes.\\

Parameters investigated were the polarizations, frequencies and intensities of the three laser beams.  The values of these parameters used in the final experiments are shown in table~\ref{paramstable}.  The beam intensities in the table are given in terms of the saturation intensity $I_S$ where 
\begin{equation}\label{satinteqn}
I_S=\frac{4\pi^2hc\Gamma}{3\lambda^3}
\end{equation}
and $\Gamma$ is the FWHM of the transition in Hz.  The detunings of the beams are referred to the separations of the centres of gravity of the \lev{S}{1/2}, \lev{P}{1/2} and \lev{D}{3/2} levels (see figure~\ref{Ca43lambda}).\\

\begin{table}
\caption{Laser parameters for the dark-resonance cooling scheme and for obtaining high fluorescence.  The intensities, detunings and their uncertainties are derived from the fit to the fluorescence spectrum in fig.~\ref{866fit}.  Polarizations are the nominal experimental values and refer to the fractional energy in the component.  The detunings and the saturation intensity $I_S$ are defined in the text.  The detunings give the position of the dark resonance; for the cooling experiments the 866\,nm beam was varied from this position to lower frequency over a range of a few MHz (see fig.~\ref{datatheory}).  Uncertainties in the detunings are correlated since the frequency separation between beams 1 and 2 is set by the EOM and that between beams 2 and 3 is determined by the dark resonance condition.}
\begin{indented}
\lineup
\item[]\begin{tabular}{@{}lllllllll}
\br
&&\centre{2}{Dark-resonance cooling}&\centre{2}{High fluorescence}&\centre{3}{Polarization}\\
\ns
&&\crule{2}&\crule{2}&\crule{3}\\
&&Intensity&Detuning&Intensity&Detuning&&&\\
&&$(I_S)$&(MHz)&$(I_S)$&(MHz)&$\pi$&$\sigma^-$&$\sigma^+$\\
\mr
397\,nm:&Beam 1&1.05(20)&1042(10)&15.8(30)&1042(10)&1/3&0&2/3\\
&Beam 2&1.24(20)&\ $-$1888(10)&18.6(30)&\ $-$1888(10)&1/3&0&2/3\\
866\,nm:&Beam 3&161(20)&\ $-$159(10)&1330(170)&201(10)&0&1/2&1/2\\
\br
\end{tabular}
\end{indented}
\label{paramstable}
\end{table}

We chose these parameters on the basis of experiment, guided by calculations using the Optical Bloch Equations (OBE).  The OBE govern the time-development of the density matrix describing the system when subject to the various radiation fields.  To specify the internal state of the system completely at any instant requires 4096 quantities, of which 64 are the state populations and 4032 coherences.  The fact that the latter are complex does not increase the number of variables because pairs of coherences involving the same states are complex conjugates of each other.  \\

To calculate $R/R'$ the time-dependent equations were required.  The state populations are never in equilibrium with the radiation field, because of the slow decay from 4P to 3D.  The effect is large enough to affect the result for $R'$ significantly.  There are even conditions under which the time-lag causes the sign of $R'$ to change.  A critical discussion of this Òdynamic effectÓ and other aspects of the theory is given in~\cite{Hugo15}.  The time-dependent equations were also needed to check that the state populations could evolve quickly enough from an arbitrary initial distribution for a given scheme to be practicable.  Speed is important in QIP processes, not only to minimize heating due to mechanisms not fundamental to Doppler cooling, but also to keep decoherence to a minimum.\\

\begin{sidewaysfigure}
 \centering
 \vspace{170 mm}
 \includegraphics[width=\textwidth]{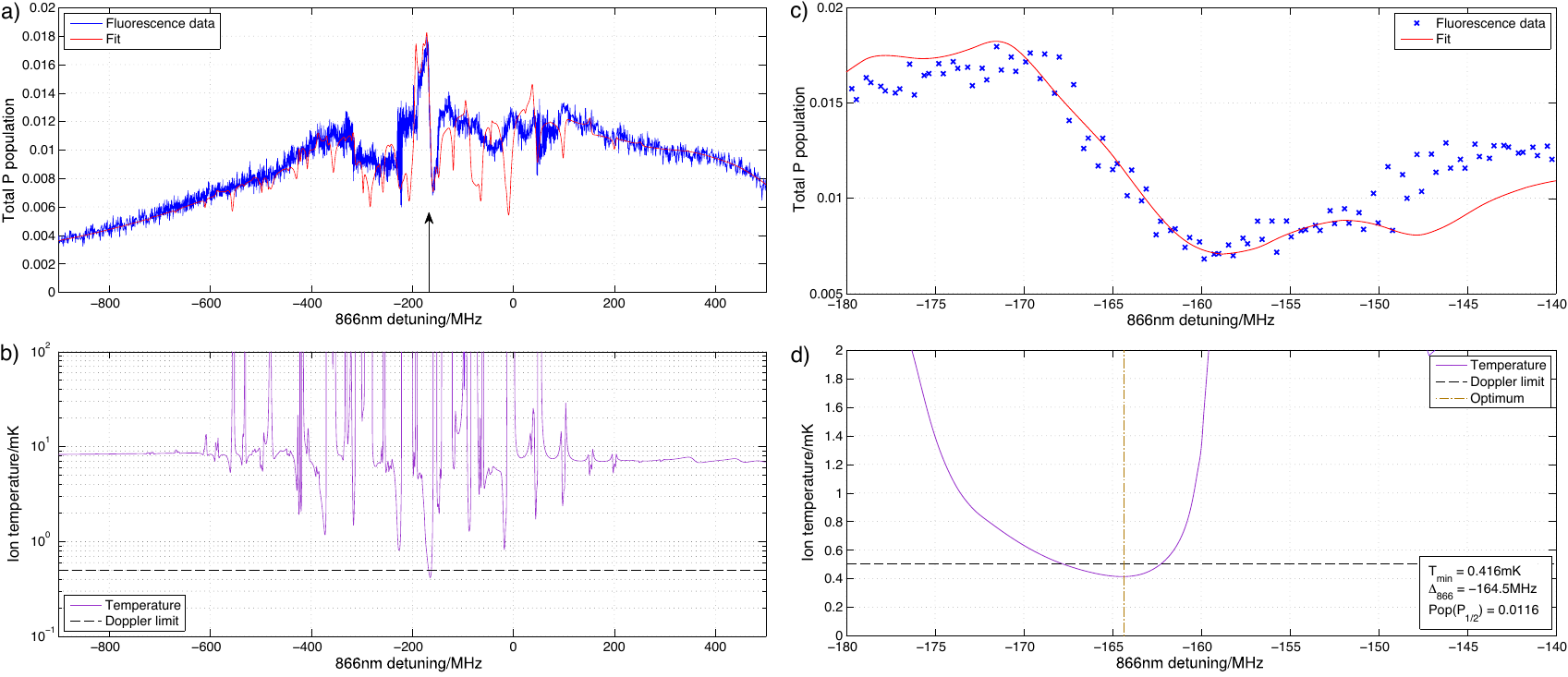}
 \caption{(a) A fitted experimental scan.  The 397\,nm fluorescence is measured as the frequency of the 866\,nm beam is scanned, and converted to total P population using the known photon counting detection efficiency.  The fit to the data, which gives the parameters shown in table~\ref{paramstable}, is based on the response of a stationary ion.  In some spectral regions, particularly on the high frequency sides of dark resonances, the observed fluorescence is significantly modified by ion heating; these regions are omitted from the fit.  The dark resonance used for cooling is at $-$159\,MHz indicated by the arrow.  (b) A plot of predicted temperature based on the fitted parameters. (c) and (d) show a close-up of the region around the dark resonance used for cooling.  At higher frequencies, the distortion of the experimental curve (c) due to heating is evident.  The orange dashed line on (d) shows the value of the 866nm detuning for which the temperature is minimal.}
 \label{866fit}
\end{sidewaysfigure}

The time-independent equations were sufficient for some diagnostic purposes.  To determine the detunings and intensities of the various beams in any given experimental situation the 397\,nm fluorescence was recorded as a function of the frequency of the 866\,nm beam.  Such scans were fitted using theoretical profiles generated for a stationary ion.  The fit which provided the data in table~\ref{paramstable} is shown in figure~\ref{866fit}(a), while figure~\ref{866fit}(c) shows a close-up of the region of the dark resonance used for cooling.  There are many regions of heating, particularly on the high frequency sides of dark resonances, but the fluorescence is close to that predicted for a stationary ion over enough of the scan to permit the extraction of the important parameters.  The laser line-widths can in principle be determined from the fit, but their extraction is complicated by two factors, as follows.  The spectral regions which are most affected by the frequency spread of the lasers show the most distortion due to heating, and in addition there are further processes associated with the motion of the ion which affect the observed profile in a very similar way.  Consider the fluorescence which is observed experimentally when the laser frequencies fulfil the condition for the dark resonance used for cooling.  This fluorescence is present partly because the laser beams are not monochromatic.  However, even if they were, there would remain two separate effects preventing the observed fluorescence from dropping to zero.  First, the motion of the ion would cause it to sample a range of frequencies (in its rest frame) over the integration period of a point on the scan.  Second, even at an instant at which the ion's velocity passes through zero, at which one might expect zero fluorescence, the dynamic effect referred to earlier prevents the internal states of the ion from reaching equilibrium with the radiation.  We therefore include in the Bloch equations parameters which enter the equations as a line-width would, but which take account of all these effects.  The best-fit values of these parameters were 1.5\,MHz (397\,nm) and 0.1\,MHz (866\,nm).\\

The temperature as a function of 866\,nm frequency as obtained from our theoretical model is shown in figures~\ref{866fit}(b) and~\ref{866fit}(d).  The minimum temperature for these conditions is 0.42\,mK, well below the two-state Doppler limit.  This drops only slightly, to 0.38\,mK, if one takes the same parameter set but assumes the lasers to be monochromatic.  However, while the curves correctly represent the qualitative behaviour of the temperature, particularly the minimum in the vicinity of the dark resonance, the theory is still approximate and one cannot expect the numerical values to be determined as reliably as the intensities and detunings.  We describe in section~\ref{experimental} how the temperatures were measured experimentally.\\
 
\begin{figure}
 \centering
 \includegraphics[width=0.8\textwidth]{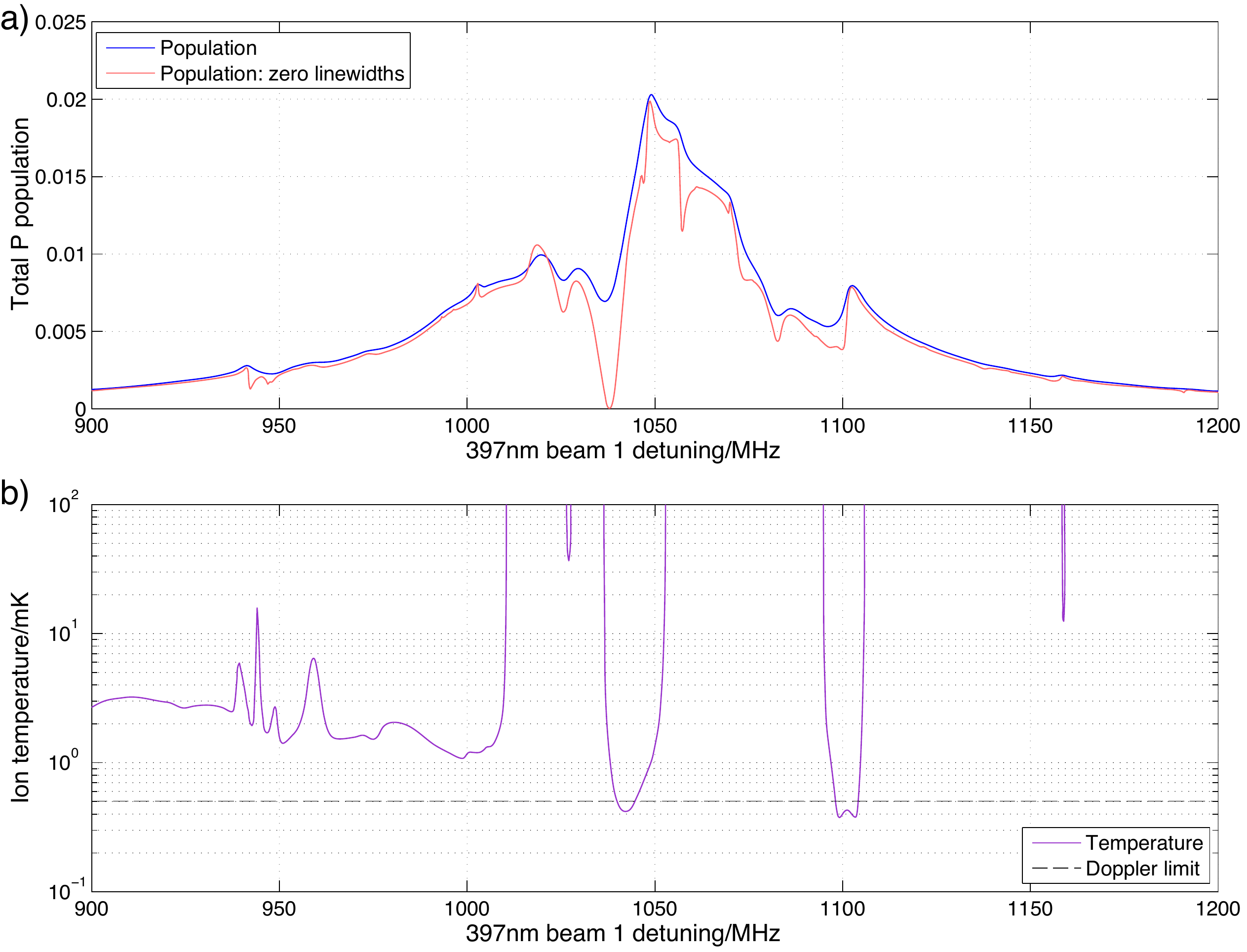}
 \caption{(a) Simulations showing the total population of the 4\lev{P}{1/2} level as a function of the frequency of beams 1 and 2 from the 397\,nm laser.  The beam 1 frequency is given on the graph axis and a constant separation of $2930$\,MHz was maintained between beams 1 and 2.  The other conditions are as given in table~\ref{paramstable}.  The blue curve incorporates the laser line-widths as determined from the fit in figure~\ref{866fit}; in the red curve, the laser line-widths have been set to zero to show that in such a case at the centre of the dark resonance used for cooling the population drops to zero. (b) The predicted temperature with the fitted laser line-widths.}
 \label{397theoryplot}
\end{figure}

To explain our choice of detunings, we show in figure~\ref{397theoryplot}(a) the total population of the P levels for a stationary ion as a function of the frequency of beam 1, as obtained from the OBE.  The predicted temperature is shown below in figure~\ref{397theoryplot}(b).  The dark resonance used for cooling is at $1042$\,MHz.  The 866\,nm detuning and the intensities of the beams are as in table~\ref{paramstable}.  Two curves are plotted, one assuming zero laser line-widths; this demonstrates that, for a stationary ion, the population drops to zero at the dark resonance.  The detuning chosen for beam 3 determines the position of the dark resonance with respect to the peak of the response curve.   In fact, our theoretical work showed that the lowest achievable temperature of the ion drops as one moves the resonance closer to and even beyond the peak.  However, it must be on the low-frequency side for the preliminary Doppler cooling, which does not involve the dark resonance, to take place.  When the ion is hot, the large and oscillating Doppler shifts cause it to sample a wide region of the response curve, and it will only cool if the net effect over a period of its motion reduces its energy.   The plot in figure~\ref{397theoryplot}(b) shows that cooling occurs over a wide range of 397\,nm detuning on the red side of the peak, whereas at higher frequencies the sharp cooling features around dark resonances are not enough to counterbalance the strong heating which occurs elsewhere.  For this reason scans of the 866\,nm frequency were a much more powerful diagnostic of experimental conditions than those of 397\,nm. \\

After some experimental tests the dark resonance was positioned at the point shown, which corresponds to a beam 3 detuning of 24MHz to the red from the \ket{3}$-$\ket{2} transition itself.  Reducing this by a few MHz then gives the required conditions for cooling.  The detuning of the 397\,nm beam 2 is not critical from the point of view of the temperature, but it does influence the overall P population and hence the level of fluorescence.  The optimum separation of beams 1 and 2 was found theoretically to be 2.930\,GHz, as in table~\ref{paramstable}, and this has been maintained over the entire scan of figure~\ref{397theoryplot}. \\

The temperature attainable, being determined by the shape of the dark resonance, varies with the beam intensities.  For 866\,nm, theory predicts that there is an optimum at about the value given in table~\ref{paramstable}.  For the two 397\,nm beams, theory suggests roughly equal intensities, predicting a decrease in temperature with decreasing intensity.  However, in practice at very low intensities heating mechanisms other than those intrinsic to the cooling process cease to be negligible and the values in the table were chosen as a compromise. Experimental studies of the variations in temperature with the intensities and detunings are described in the next section.\\

The polarization states of the lasers are not critical, provided that they meet the conditions given earlier.  Our experimental conditions made it convenient to linearly polarize beam 3 perpendicular to the B-field direction, so that the energy was equally divided between $\sigma^-$ and $\sigma^+$ components, while the energy in the 397\,nm beams was split in the ratio 2:1 between $\sigma^+$ and $\pi$.   \\

\section{Experimental details and results}
\label{experimental}

\subsection{Experimental System}

The general method was as follows.  The ion was Doppler-cooled for a few milliseconds, then further sideband-cooled to the ground state making use of the Raman Zeeman transition at 57.480\,MHz between states \lev{S}{1/2} \ket{F=4, M = +4} (that is, \ket{1} in the cooling cycle) and \lev{S}{1/2} \ket{F=4, M = +3}.  We now denote these states \ket{4, +4} and \ket{4, +3} respectively.  At any stage the temperature of the ion could be measured by sideband thermometry~\cite{Leibfried03} on the same transition.\\

We perform these experiments in the same surface-electrode trap used to demonstrate the intermediate-field qubit \cite{Harty14}; it is described in detail elsewhere \cite{Allcock13}.  The mode frequencies were 477\,kHz ($z$, axial), 3270\,kHz ($x$, radial) and 3630\,kHz ($y$, radial). Unless stated otherwise, all sideband thermometry is on the $x$ mode.  From the measured Lamb-Dicke parameters (see section~\ref{gscool}) we find that this mode is oriented at 40(2)$^\circ$ to the trap surface.\\

\begin{figure}
 \centering
 \includegraphics[width=0.65\textwidth]{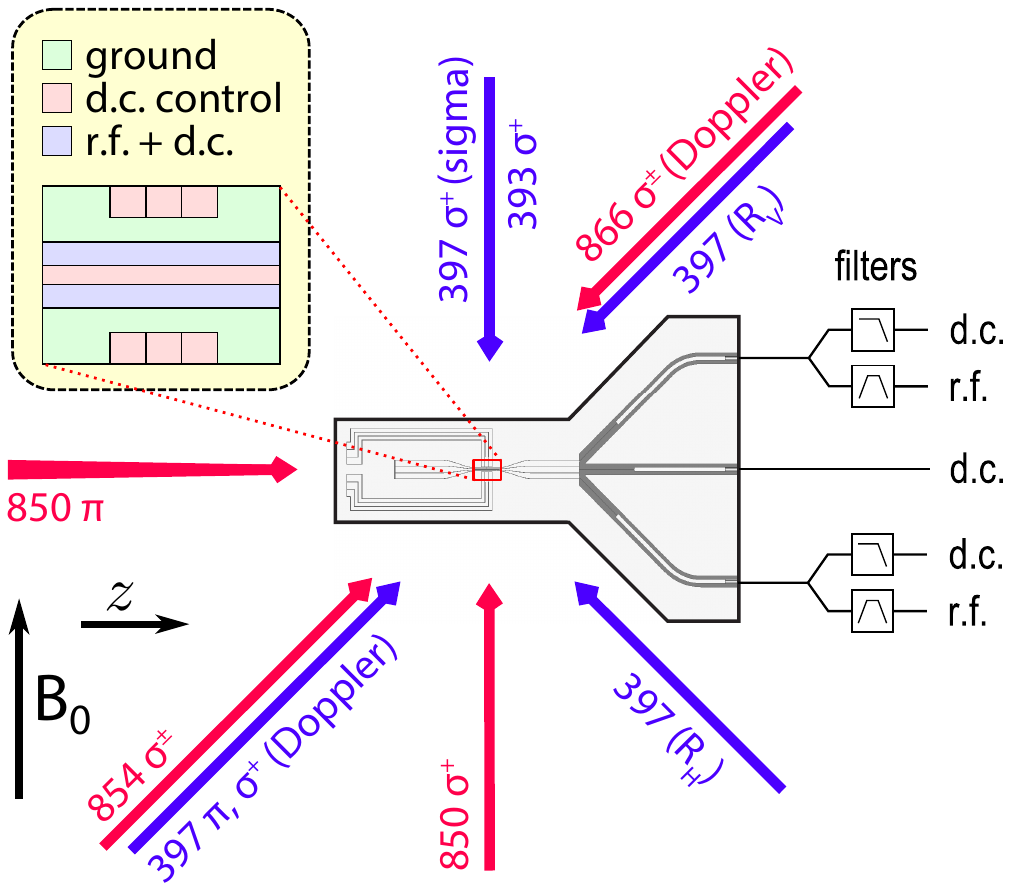}
 \caption{Diagram showing the laser beam orientations and polarizations, and (inset) the central electrode layout of the surface trap. Radio-frequency (r.f., 40MHz) and d.c. voltages for trapping are combined using filters as indicated.  The 397\,nm Doppler-cooling beam is elliptically polarized such that it contains only $\sigma^+$ and $\pi$ polarizations, with intensity ratio $I_{\sigma^+}/I_\pi=2$, and counter-propagates along the linearly ($\sigma^\pm$) polarized 866\,nm Doppler-cooling beam. Circularly ($\sigma^+$) and linearly ($\pi$) polarized beams are used for Raman repumping and state readout (the 850nm $\pi$ beam is at $\sim 45^\circ$ to the plane of the figure and reflects off the trap surface). The linearly-polarized Raman beams $R_V$ and $R_H$ are used for Raman sideband cooling and thermometry of the radial motion of the ion. Photoionization beams (not shown) for loading the trap copropagate with the 397\,nm Doppler beam. The quantization axis is defined by a static magnetic field $B_0=146$\,G. The trap axial direction is denoted by $z$; the radial $x$ mode lies at 40(2)$^\circ$ to the plane of the electrodes.}
 \label{Beams}
\end{figure}

Figure~\ref{Beams} shows the beam geometry used.  In addition to the 397\,nm and 866\,nm beams required for Doppler cooling, there are three further beams at 397\,nm.  One of these, with $\sigma^+$ polarization, ensures that the ion is optically pumped to the state \ket{4,+4} after the Doppler cooling stage.  The others are the Raman beams used to induce the $\ket{4,+4} \Leftrightarrow \ket{4,+3}$ transition.  There are also beams at 393\,nm, 850\,nm and 854\,nm, used for qubit shelving readout, see~\cite{Harty14}.  Single-mode fibres deliver each beam from the laser system to the trap.  Each beam has a photodiode after the fibre which is used with an FPGA-based intensity stabilization system.  This system allows for different power set points to be used at different points in an experimental sequence.  Beams that share the same path are combined on dichroic mirrors except the 393\,nm and 397\,nm $\sigma^+$ which are combined before the fibre.\\  

The two 397\,nm Doppler cooling beams are derived from the same laser, the frequency separation being provided by an EOM operating at 2.930\,GHz.  Beam 2 is the carrier, beam 1 the first higher-frequency sideband.  The microwave power is adjusted so as to give this sideband $85(10)\%$ of the intensity of the carrier.  The first side-band to lower frequency, and all higher orders, are too far detuned to produce any significant effect on the ion.  The polarizations of the beams were set as in table~\ref{paramstable} using standard optical components (see~\cite{HartyTh} for a detailed description).\\

The two 397\,nm Raman beams $R_H$ and $R_V$, where the suffixes denote horizontal and vertical linear polarization, are obtained from a single laser, detuned from the transition \lev{S}{1/2} $-$ \lev{P}{1/2} by $\Delta\approx-30$\,GHz.  The Rabi frequency on the $\ket{4,+4} - \ket{4,+3}$ carrier transition is 171(2)\,kHz.  $R_V$ counter-propagates with the 397\,nm Doppler cooling beams, while $R_H$ is perpendicular to $R_V$ such that there is no coupling to the axial modes.  Each beam passes through its own switching AOM.  

\subsection{Temperature optimisation and measurement}
\label{expopt}

\begin{figure}
 \centering
 \includegraphics[width=0.75\textwidth]{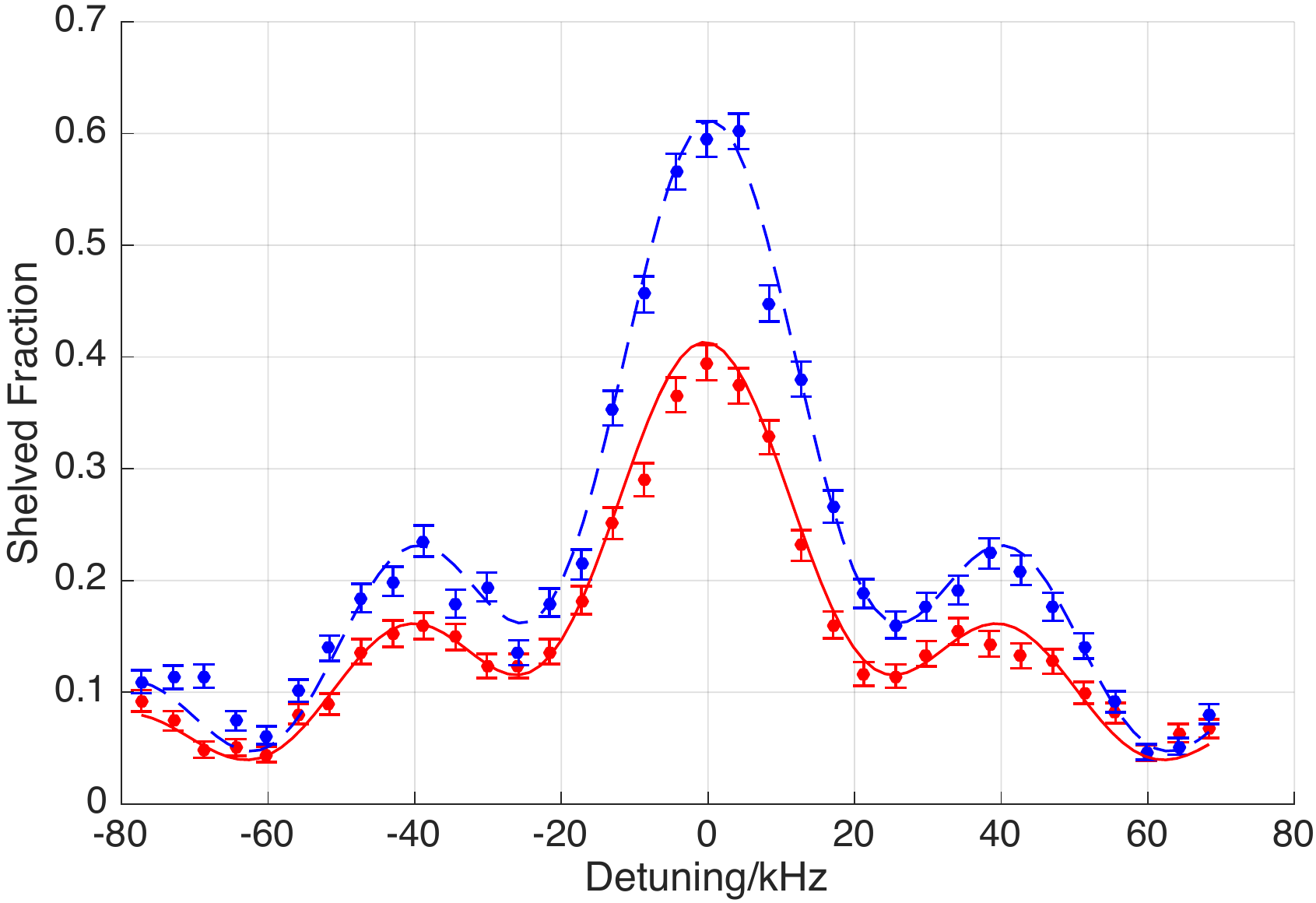}
 \caption{Red and blue (dashed) Raman first sideband resonances with our optimised Doppler cooling parameters. A $30\,\mu$s Raman probe pulse was used and the fit gives $\bar{n}_x$= 1.97(5). }
 \label{DopplerSB}
\end{figure}

To optimise the cooling parameters, we tuned the 397\,nm and 866\,nm powers and detunings while looking at the height of the second red Raman sideband. Our pulse length was $\sim$50\,$\mu$s (significantly shorter than the second sideband $\pi$-time). We minimised the spin-flip probability to minimise the temperature.  The best parameters we found were 5\,$\mu$W of 397\,nm (total beam power at trap) and 40\,$\mu$W of 866\,nm, with the 397\,nm near resonance, giving the analysed parameters in table~\ref{paramstable}.  In these experiments, 6.7\,ms of Doppler cooling was followed by 200\,$\mu$s of optical pumping to prepare $\ket{4,+4}$.  After optimising the parameters, we took a pair of red and blue first Raman sideband scans to measure the temperature (fig.~\ref{DopplerSB}). 

\begin{sidewaysfigure}
 \centering
 \vspace{160 mm}
 \includegraphics[width=0.8\textwidth]{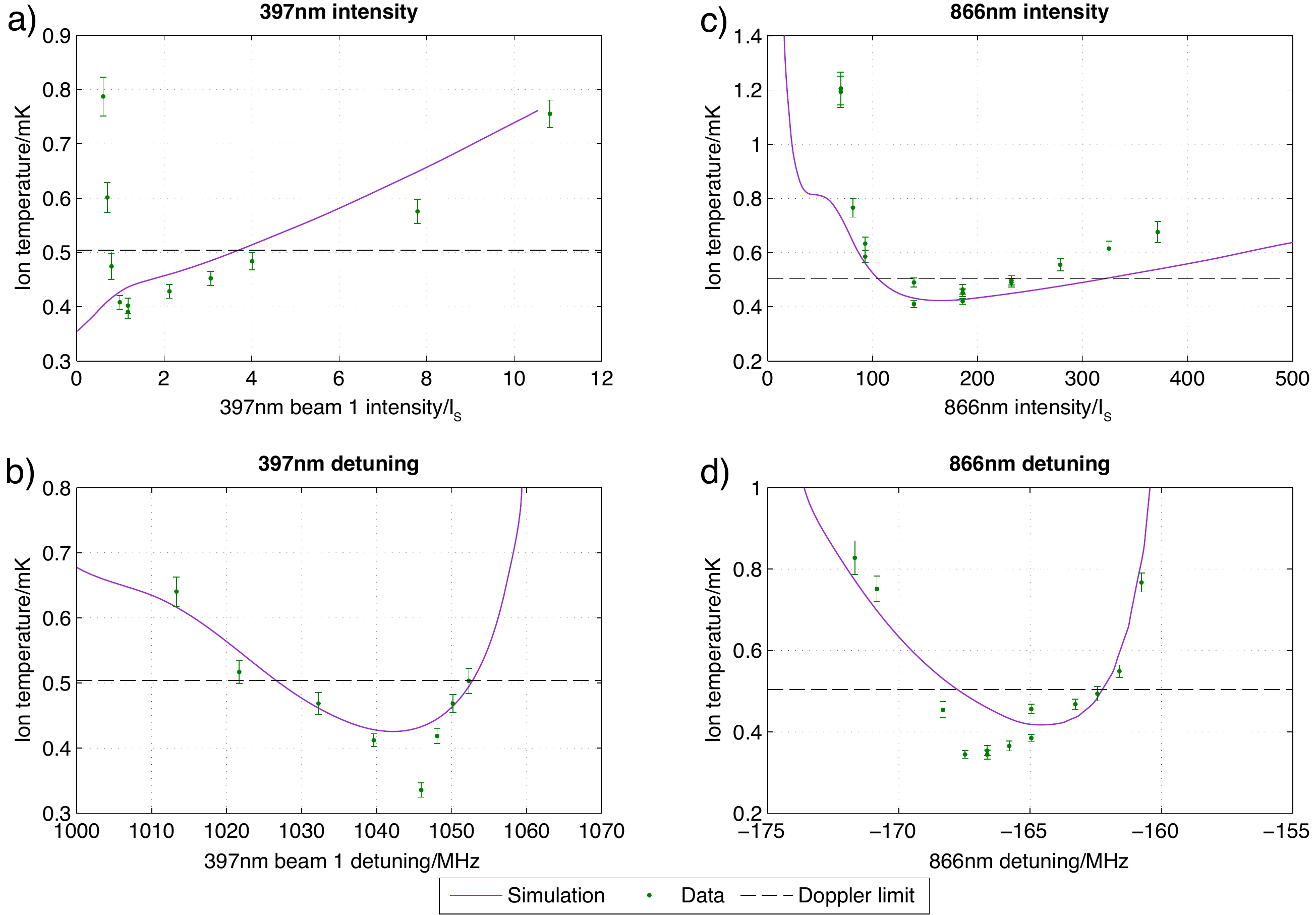}
 \caption{Experimental and simulated temperature as a function of: (a) intensity in 397\,nm beam 1 (beam 2 maintained at a constant ratio), (b) detuning of 397\,nm beam 1 (beam 2 maintained at a constant separation of $-2930$\,MHz), (c) power in the 866\,nm beam, (d) detuning of the 866\,nm beam.  All parameters apart from the scanned parameter are kept at their optimised values except for (b) where the 866\,nm detuning was re-optimised at each 397\,nm detuning.  The dashed lines show the 2-level Doppler temperature limit for comparison.  The temperature error bars are the statistical uncertainty from the sideband fits.  The main systematic error in this data comes from changes in detuning caused by drifts in the lasers' frequency reference cavities. These were estimated to be below 2\,MHz over the time taken to obtain the data for each of the graphs.}
 \label{datatheory}
\end{sidewaysfigure}

We investigated the dependence of the temperature reached by Doppler cooling on the 397\,nm and 866\,nm powers and detunings. For the 397\,nm and 866\,nm powers as well as the 866\,nm detuning, we scanned one parameter (power/detuning) at a time, keeping all others at their optimised values. At each value of the scanned parameter, we took a red and blue first sideband scan, which we fitted to determine the temperature. For the 397\,nm detuning scan we re-optimised (using the second red sideband) the 866\,nm detuning at each 397\,nm detuning before taking the red and blue first sideband scans. This was to separate the question of how close to resonance the 397\,nm should be from that of where on the dark resonance one ideally wants to be (which we had already investigated by looking at the 866\,nm detuning). The results of these scans are shown in figure~\ref{datatheory}.  The features of the experimental data are satisfactorily reproduced by the theory, apart from the rise in temperature at very low 397\,nm intensity which we attribute to the neglect of heating mechanisms other than those intrinsic to the Doppler cooling process. 

\subsection{High fluorescence for readout}

To switch between sub-Doppler and high-fluorescence cooling we use the other set of laser parameters in table~\ref{paramstable}.  This increases the 397\,nm power to 75\,$\mu$W (keeping its detuning constant), as well as increasing the 866\,nm power to 330\,$\mu$W and increasing the 866\,nm frequency by 360\,MHz (this requires a separate AOM).  We then obtain a fluorescence count rate comparable to that from a single, saturated, $^{40}$Ca$^+$ ion, at $\approx25,000$\,s$^{-1}$ with a net photon detection efficiency of 0.17\% (11\% P-state population), which is sufficient to allow readout fidelities of $\approx99.98\%$.

\subsection{Ground state cooling}
\label{gscool}

We used a continuous Raman sideband scheme to cool the ion to the ground state.  In this scheme the 397\,nm $\sigma^+$ and 866\,nm repumpers are applied at the same time as driving the red sideband transition with the Raman lasers.  The full experimental sequence was:

\begin{enumerate}[1]
\item Doppler cool on dark resonance for 2.5 ms
\item Continuous cooling on 1st red sideband of the $y$-mode for 0.5\,ms
\item Continuous cooling on 1st red sideband of the $x$-mode for 1.5\,ms
\item Continuous cooling on 1st red sideband of the $y$-mode for 0.5\,ms
\item Continuous cooling on 1st red sideband of the $x$-mode for 1\,ms
\item Optical pumping to $\ket{4,+4}$
\item Thermometry by comparison of the first red and blue Raman sidebands.
\end{enumerate}

\begin{figure}
 \centering
 \includegraphics[width=0.70\textwidth]{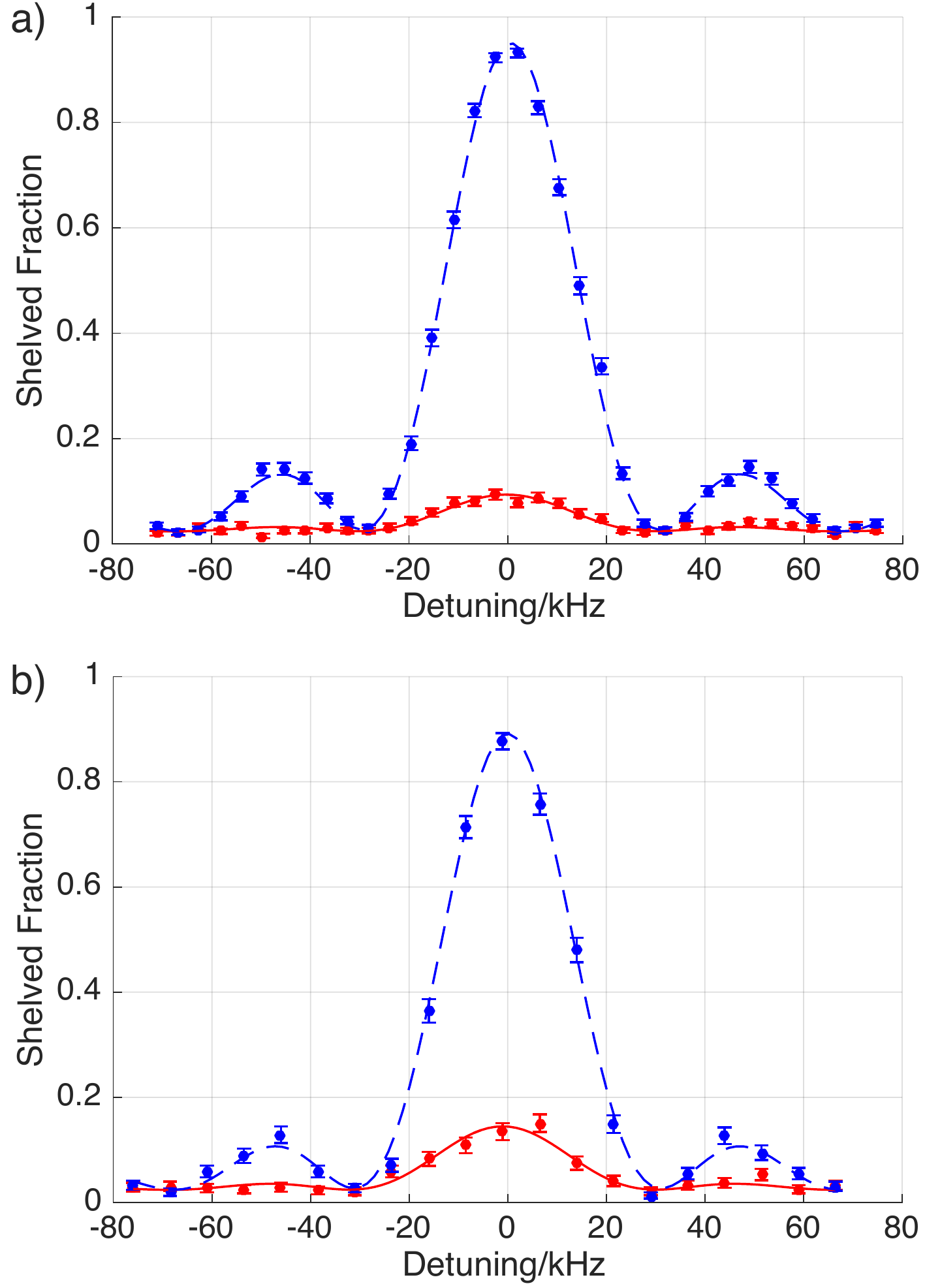}
 \caption{Red and blue (dashed) first sideband detuning scans of the $x$ (a) and $y$ (b) radial modes. Fits give $\eta_x$ = 0.098 (2), $\bar{n}_x$ = 0.08 (1), $\eta_y$ = 0.082 (2), $\bar{n}_y$ = 0.16 (2), where $\eta$ is the corresponding Lamb-Dicke parameter.}
 \label{gsfig}
\end{figure}

Using this scheme we reached $\bar{n}_x$ = 0.08(1) and $\bar{n}_y$ = 0.16(2), as shown in fig.~\ref{gsfig}.  With larger Raman power and detuning $\Delta$ and more careful optimisation it is expected that even lower temperatures could be achieved, but this temperature is sufficient for implementing 2-qubit microwave or laser-driven gates with good fidelity.

\section{Conclusion}

We have demonstrated that despite the difficulties introduced by the large number of widely-separated states in the \Ca{43} ion when subject to a field of 146\,Gauss, it is possible to cool it to around the Doppler limit using just 3 laser frequencies by making use of a dark resonance.  To achieve this requires the fluorescence to be kept low, because spontaneous emission is partly responsible for the heating experienced by the ion during Doppler cooling.  However, the conditions are readily adaptable to allow the fluorescence of the ion to be raised so that such operations as qubit read-out can be carried out with the necessary precision.  Furthermore we have shown that further application of sideband cooling allows preparation of both radial modes in the ground state.  Experiments which exploit the techniques described to prepare ions for two-qubit entangling gates are in progress \cite{Harty15}.  Simulations~\cite{HugoTh} indicate these techniques will also be suitable for use in \Ca{43} at 288\,Gauss (the next highest field clock qubit beyond 146\,Gauss) and for the yet to be investigated clock qubits of \Sr{87} and \Ba{135, 137}.

\ack
The authors thank L. Guidoni for assistance in setting up the experiment, N. M. Linke for producing figure 5 and D. J. Szwer for assisting with preliminary simulations.  TPH gratefully acknowledges support from St. John's College, Oxford.  This work was supported by an EPSRC Science \& Innovation award.

\clearpage

\section*{References}

\bibliographystyle{unsrt}
\bibliography{Coolpaperarxiv}

\end{document}